\documentclass[%
preprint,
 amsmath,amssymb,
 aps,
 prb,
]{revtex4-1}  %if pra is needed, plz remove the %
\usepackage{graphicx}% Include figure files
\usepackage{dcolumn}% Align table columns on decimal point
\usepackage{bm}% bold math
\usepackage{epstopdf}
\setcitestyle{super}

\preprint{APS/123-QED}

\usepackage{color}
\usepackage{bbold}
\usepackage{tabularx, booktabs}
\usepackage[normalem]{ulem}
\newcolumntype{Y}{>{\centering\arraybackslash}X}
\newcolumntype{Z}{>{\hsize=1.1\hsize\centering\arraybackslash}X}

\newcommand*{\LAOSTO}{LaAlO$_3$$\slash$SrTiO$_3$}
\newcommand*{\LAO}{LaAlO$_3$}
\newcommand*{\STO}{SrTiO$_3$}
\newcommand*{\SMO}{SrMnO$_3$}
\newcommand*{\LMO}{LaMnO$_3$}
\begin{document}

\title[First-principles design of ferromagnetic monolayer MnO$_2$ at the complex interface]{First-principles design of ferromagnetic monolayer MnO$_2$ at the complex interface}

\author{Rui-Qi Wang$^{1,2}$}
\author{Tian-Min Lei$^{1}$}
\author{Yue-Wen Fang$^{3,4,5}$}
\email{fyuewen@gmail.com}

\affiliation{
$^1$ School of Advanced Materials and Nanotechnology, XiDian University, Xi'an 710126, China \\
$^2$ School of Electronic Engineering, Xi'an Aeronautical Institute, Xi'an 710077, China \\
$^3$ Key Laboratory of Polar Materials and Devices (MOE), Ministry of Education, Department of Electronics, East China Normal University, Shanghai 200241, China \\
$^4$ Centro de F\'{i}sica de Materiales (CSIC-UPV/EHU), Manuel de Lardizabal Pasealekua 5, 20018 Donostia/San Sebasti\'{a}n, Spain \\
$^5$ Fisika Aplikatua Saila, Gipuzkoako Ingeniaritza Eskola, University of the Basque Country (UPV/EHU), Europa Plaza 1, 20018 Donostia/San Sebasti\'{a}n, Spain
}

\vspace{10pt}

\date{\today}
\begin{abstract}
 Rapidly increasing interest in low-dimensional materials is driven by the emerging requirement to develop nanoscale 
  solid-state devices with novel functional properties that are not available in three-dimensional bulk phases. 
  Among the well-known low-dimensional systems, complex transition metal oxide interface holds promise for broad
   applications in electronic and spintronics devices. Herein, intriguing metal-insulator and 
   ferromagnetic-antiferromagnetic transitions are achieved in monolayer MnO$_2$ that is sandwiched into 
   SrTiO$_3$-based heterointerface systems through interface engineering. 
   By using first-principles calculations, we modeled three types of SrTiO$_3$-based heterointerface systems with different interface terminations and performed a comparative study on the spin-dependent magnetic and electronic properties that are established in the confined MnO$_2$ monolayer. First-principles study predicts that metal-insulator transition and magnetic transition in the monolayer MnO$_2$ are independent on the thickness of capping layers. Moreover, 100$\%$ spin-polarized two-dimensional electron gases accompanied by robust room temperature magnetism are uncovered in the monolayer MnO$_2$. Not only is the buried MnO$_2$ monolayer a new interface phase of fundamental physical interest, but it is also a promising candidate material for nanoscale spintronics applications. Our study suggests interface engineering at complex oxide interfaces is an alternative approach to designing high-performance two-dimensional materials.
\end{abstract}
\maketitle

\section{Introduction}

A central goal in materials physics and materials engineering is to achieve precise control over low-dimensional materials
 at the atomic level with the reduction of the size of solid-state devices~\cite{Yang2023Science,Wu2022-JPCM,Shen2019_2DFJT,Tong2016CompMaterSCI}. 
 2D CrI$_3$ has a measured Curie temperature of 45 K, which has been experimentally synthesized, and stimulated the active research in 2D ferromagnetic materials~\cite{Huang2017Nature-2D-FM}{} because they are more suitable for the application of nanoscale devices than the traditional magnetic materials such as diluted magnetic semiconductor ZnO\cite{AbdelBaset2016}. 
 For example, the ferromagnetic order persisting down to the bilayer limit, and room temperature magnetic order have been demonstrated respectively in 2D Cr$_2$Ge$_2$Te$_6$\cite{CGong2017Nature-2D-FM}{} and monolayer VSe$_2$\cite{Bonilla2018}{}. Very recent studies have predicted magnetic ordering in a variety of 2D materials such as CrX$_3$ (X = I, Br, Cl)\cite{Liu2021-Nanoscale}{}, NiI$_2$\cite{Song2022-Nature}{}, Cr$_2$Te$_3$\cite{Zhong2021-Nano}{}, and  Fe$_3$GeTe$_2$\cite{Fei2018-Natmat}{}. However, only a few two-dimensional materials own robust magnetism above room temperature\cite{Song2021-APL,Zhang2021-Nat}{}.

One alternative way to incorporate magnetism into two-dimensional materials is to synthesize transition metal oxide nanosheet. Monolayer MnO$_2$ is a 2D layered semiconducting transition metal oxide material that has been reliably experimentally synthesized and studied extensively with computational
methods\cite{Kitchaev2016,Rong2019,Henry2016-MnO2-AdvMater}{}. Kan et al. predicted a possibly stable graphene-like antiferromagnetic 
MnO nanosheet and claimed single-layer doped MnO could become a half-metallic ferromagnet with Curie temperature of 
350 K\cite{Kan2013}{}. Diffusion Monte Carlo (DMC) and DFT+$U$ methods are used to calculate the magnetic properties of monolayer MnO$_2$ and find that the ferromagnetic ordering is more favorable than antiferromagnetic one\cite{Wines2022}{}. Regardless of the difficulty in preparing free standing MnO nanosheet with graphene-like structure,
some recent studies have succeeded in manufacturing transition metal oxide nanosheets. For example, we have demonstrated a direct metallic conversion from nickel hydroxide nanosheets to nickel metal nanostructures by thermal annealing in vacuum, and proved that the converted nickel metallic structures exhibit ferromagnetic behavior revealed by x-ray magnetic circular dichroism measurement\cite{Naruo2020-Nanotechnology}{}. In addition, a method of wet-chemical synthesis was recommended to make two-dimensional transition metal nanosheets\cite{Tan2015}{}, and this method has been applied to 
make $\alpha$-Fe$_2$O$_3$---a magnetic semiconductor with intrinsic ferromagnetism at room temperature\cite{Cheng}{}. Very recently, a variety of studies on the synthesis of magnetic MnO$_2$ nanorods\cite{Gangwar2021}{} and Fe$_3$O$_4$/MnO$_2$ nanocomposite with their applications as sorbents\cite{Chen2020}{}, catalysts\cite{Bakhtiarzadeh2021}{}, and photodegradation agents\cite{Dubey2021}{} have been published. Moreover, MnO$_2$ nanoparticles and its nanocomposite with nitrogen-doped graphene have been fabricated via simple hydrothermal synthesis procedure using water as a solvent, in which the strong ferromagnetic character of nanohybrid helps in easy separation of catalyst even with a bar magnet\cite{Singh2019-MOLECULAR-PHYSICS}{}. Nevertheless, using the above methods, defects could be introduced to two-dimensional nanosheets due to the sensitive 
reactions to chemical circumstances (e.g., aqueous solution), which not only makes it difficult to realize flexibly 
artificial control of properties of materials, but also may have a destructive effect on the spin ordering\cite{Awad2018-J-MATER-SCI-MATER-EL,Shah2019-J-MATER-SCI-MATER-EL}{}. 
A better alternative to obtain controllable couplings of lattice, charge and spin order in defect-free materials 
is to make use of the appealing properties of two-dimensional transition metal oxide layers sandwiched at the 
complex perovskite oxide interfaces. Even so, little attention has been paid to this kind of buried two-dimensional materials up till now.

Benefiting from the steady development of solid-state synthesis method (e.g., pulsed laser deposition), 
great success has been achieved in preparing defect-free complex oxide interfaces in the latest decade.  One of the most representative instances is the transition metal complex oxide \LAOSTO{} interface confining quasi two-dimensional electron gas (2DEG) at the nanoscale\cite{Han-Fang2015,PhysRevB.93.214427}{}. Control variables have been successfully utilized to manipulate the 2DEG, such as fully optical modulation\cite{Niu2022}{}, electronic reconstructions\cite{Fang2022}{}, ferroelectric polarization\cite{Sharma2015}{}, electric field-effect through the gate\cite{Monteiro2017}{}. Surprisingly, the fascinating interface even also supports superconductivity\cite{Han.APL.2014}{}, electronic
phase separation\cite{Ariando.2011,Scopigno2016}{}, and strong Rashba spin-orbital coupling\cite{Herranz.Nat.Comm.2015}{}. The structure of \LAOSTO{} interface is actually an atomic stacking sequence of perovskite structure ABO$_3$. 
 This scenario of stacking is very popular to explore interesting properties at the two-dimensional perovskite 
 materials\cite{Binglun2015,Jeong2014,Tsymbal2015}{}. For ferromagnet
SrRuO$_3$ below 2-4 unit cells, ferromagnetic/metal to antiferromagnetic/insulator transitions were predicted as the result of inversion symmetry breaking combined with non-degenerate Ru $4d$ orbitals via crystal field splitting\cite{Chang2009,Huang2018,Liu2021-appliedmatertoday}{}.

Herein, intrigued by the emergent phenomenon of stacking perovskite oxides, the aim of the present article is 
to theoretically design new two-dimensional transition metal oxides that can be accessible by solid-state synthesis methods.
 By carrying out first-principles calculations, metal-insulator transition involving antiferromagnetism-magnetism 
 transition is observed in the monolayer MnO$_2$, and 100$\%$ spin-polarized two-dimensional electron gas is 
 extremely confined into the ferromagnetic monolayer with quite robust ferromagnetism with at least 3.11 
 $\mu_B$$\slash$Mn. The magnetism is predicted to remain at room$\slash$higher temperatures within mean-field
  theory and Heisenberg model.

\section{Method}
First-principles density functional theory calculations are all performed using the Vienna $ab${} $initio${} 
Simulation Package (VASP)\cite{Kresse1996,Kresse-PRB-1996} along with the projector augmented wave method~\cite{Blochl1994}
and local density approximation (LDA). The kinetic energy cutoff of 500 eV is used for expanding the
plane-wave basis set. $\Gamma$ centered $7 \times 7 \times 1$ and $14 \times 14 \times 1$ Monkhorst-Pack
 \emph{k}-meshes\cite{Monkhorst1976} are used for total energy calculations and densities of states 
 calculations, respectively.

The on-site repulsion (Hubbard $U$) item in the approach of Dudarev et al.~\cite{Dudarev-LDAU-PRB1998} is introduced in Hamiltonian to include approximately the effects of 
localized electronic correlations that are missing from standard LDA calculations. The effective Hubbard $U$ parameters for Ti-3$d$, La-4$f$, and Mn-3$d$ orbitals are 5 eV, 11 eV and 4.5 eV, respectively. These Hubbard $U$ values have been proved in some previous study\cite{PhysRevB.93.214427,Henry2016-MnO2-AdvMater} to provide insightful interpretations of interface phenomena in \STO-based heterostructure systems.

In order to avoid a spurious electric field, we use symmetrical slab models for all the heterointerface structures.
 A vacuum spacing of at least 26 \AA{} is used in each slab model to eliminate the interactions between repeated slabs.
  Our LDA+$U$ calculation shows the estimated equilibrium lattice constant of bulk SrTiO$_3$ is 3.904 \AA{} which is
   very close to its experimental value of 3.905 \AA. Thus, the in-plane lattice constants of the slab models 
   in this study are all fixed at the optimized lattice constant of bulk \STO{}~(3.904 \AA) to simulate the epitaxial growth on the \STO{} substrates.

In the slab models, all coordinates of atomic positions along [001] direction perpendicular to the interfaces are
 fully relaxed until the forces are less than 0.01 eV$\slash$\AA{}, meanwhile the energy convergence criterion of
  10$^{-6}$ eV is guaranteed. %the \emph{\textbf{c}}-direction

\section{Results}
\subsection{Electrostatic field analysis}\label{results-sec1}
\begin{figure}
  \centering
  \includegraphics[width=\linewidth]{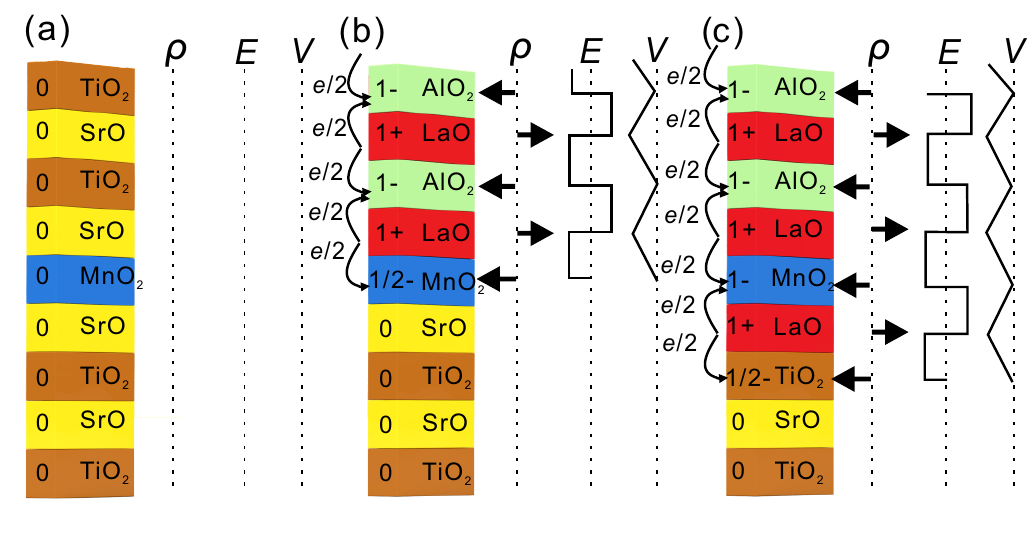}\\
  \caption{(Color online) Reconstruction of electrostatic field. (a) In case 1, there is no charge transfer at the 
  non-polar interfaces. (b) and (c) are for case 2 and case 3, respectively, in which charge transfer is expected 
  to cause the reconstruction of the electrostatic field. In case 2, electrons transfer from one side to MnO$_2$ monolayer; in case 3, electrons transfer from two sides to MnO$_2$ monolayer.
  }
  \label{fig:FIG1} %
\end{figure}

In Figure \ref{fig:FIG1}, we devise three MnO$_2$-sandwiched heterointerfaces which are named as case 1, case 2 
and case 3. In spite of the practical difficulty in preparing such complex oxide interfaces, an alternative way 
to accomplish this goal can be realized in two steps. More specifically, one can first grow a unit-cell 
of \SMO\ or \LMO\ on TiO$_2$ terminated \STO\ substrate, and then cover the MnO$_2$ monolayer terminations using nonpolar
 \STO\ or polar \LAO\ overlayers. 
 By following these steps, one former study has succeeded in fabricating such complex oxide
  heterointerfaces through layer-by-layer growth technique~\cite{Henry2016-MnO2-AdvMater}.
%More specifically, one can firstly grow an unit-cell of \SMO\ composed of two formally neutral planes or \LMO\ consisting of -1 MnO$_2$ and +1 LaO planes on TiO$_2$ terminated \STO\ substrate, and then cover the MnO$_2$ monolayers using nonpolar \STO\ or polar \LAO\ overlayers. By following these steps, we have fabricated such heterointerface samples by layer-by-layer growth technique through pulsed laser deposition in our recent study.\cite{Henry.2016}

Bulk \SMO\ is composed of two formally neutral planes, i.e. SrO and Mn$^{4+}$O$_2$ planes; while bulk \LMO\ 
is constructed by -1 Mn$^{3+}$O$_2$ and +1 LaO planes. The multivalent nature of Mn cations, which can exist 
as either Mn$^{3+}$ or Mn$^{4+}$, provides a possibility to modulate the properties of MnO$_2$ monolayers that 
are confined in \STO-based heterointerfaces. As shown in Figure \ref{fig:FIG1}, different interface terminations
 adjacent next to the MnO$_2$ monolayer correspond to three types of electrostatic boundary conditions. 
 These boundary conditions dominantly control the charge reconstruction and significantly affect the atomic
 and electronic structures of MnO$_2$ monolayer.

Formation energy of interface can be used to characterize the chemical stability of the interface, which is expressed by the following formula:
\begin{equation}
E_{\rm Formation} = E_{A/B/C}-E_{A}-E_{B}-E_{C}  
\end{equation}
where $E_{A/B/C}$, $E_A$, $E_B$ and $E_C$ refer to the total energies of the interface, and the three different bulk components.
The formation energies of interfaces are -4.44 eV/supercell, -7.34 eV/supercell, and -12.06 eV/supercell for case 1, case 2, and case 3, respectively, suggesting the interfaces are chemically stable. It is noted that the three different interfaces have been experimentally fabricated and validated our theoretical calculations of interface stability~\cite{Henry2016-MnO2-AdvMater}.

% \begin{table}%[!ht]
% \caption{The formation energies ${E_{\rm Formation}}$ (eV) of the three heterointerfaces. }
% \begin{center} \footnotesize
% \begin{tabular}{l  c c c c c c l|c|c|c|c|} \hline \hline %
% Case name & Ground state  &  ${E_{\rm Formation}}$\\
% 		\hline
%  case 1 & AFM & \ \ \ \ \ \ \ \ \ \  \ -4.44 \\
%  case 2 & FM & \ \ \ \ \ \ \ \ \ \ \  -7.34 \\
%  case 3 & FM & \ \ \ \ \ \ \ \ \ \ \ \ -12.06 \\
% \hline
% \hline
% \end{tabular}
% \end{center}
% \label{tab:groundstate}
% \end{table}

In case 1, one unit-cell of \SMO{} is sandwiched between the \STO\ substrate and \STO\ capping layers, charge
 reconstruction is  generally not expected at the nonpolar/nonpolar/nonpolar interface. While in case 2, 
 monolayer MnO$_2$ on the \STO\ substrate is buried by LaO terminated \LAO\ overlayers.
  At this nonpolar$\slash$nonpolar$\slash$polar -SrO-TiO$_2$$\slash$SrO-MnO$_2$$\slash$LaO-AlO$_2$- interface, 
  the electric potential ($V$) would diverge with the thickness of \LAO\ if there is no atomic or electronic reconstructions. 
  This divergence catastrophe can be avoided by moving half an electron to the monolayer MnO$_2$ (Figure \ref{fig:FIG1}b). 
  As a result, Mn is in the ${(4-x)+}$ valence state (0 $<$ x $\le$ 1). Case 3 is constructed by successively stacking three 
  perovskites \STO, \LMO, and \LAO. In this structural arrangement, a nonpolar$\slash$polar$\slash$polar 
  -SrO-TiO$_2$$\slash$LaO-MnO$_2$$\slash$LaO-AlO$_2$- interface is developed. Analogously to case 2, the charge at the 
  interface is reconstructed to protect the stability of the interface from potential divergence through transferring
   half an electron per two-dimensional unit-cell to TiO$_2$ layer (Figure \ref{fig:FIG1}c). The corresponding valence
    state of Mn in this case would take the value of (3-$x$)+.

Aiming to further identify the presence of electronic reconstruction with a microscopic description, and discover 
the novel properties, a different method based on the principles of quantum mechanics is indispensable. 
First-principles calculation rooted in the framework of density-functional theory,
 for instance, is a tried-and-true exemplar of the successful method to detect electronic and
  magnetic properties of materials\cite{Fang.2014} and we apply it to simulate the three heterointerfaces in our study.
\subsection{Magnetic and electronic properties}
\begin{figure}
\centering
\includegraphics[width=\linewidth]{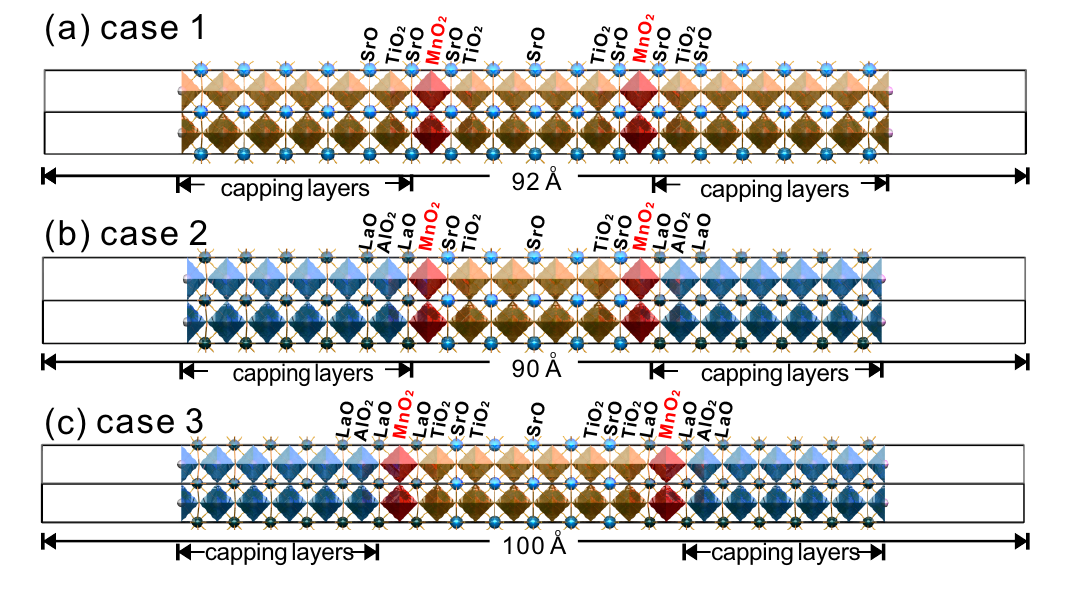}
\caption{(Color online) Relaxed structures of the three cases with 6 capping layers. (a) case 1. 
(b) case 2. (c) case 3. There is a vacuum region of at least 26 \AA{} in each case to separate the repeated slabs.
}
\label{fig:models6capping}
\end{figure}

\begin{figure}
\centering
\includegraphics[width=\linewidth]{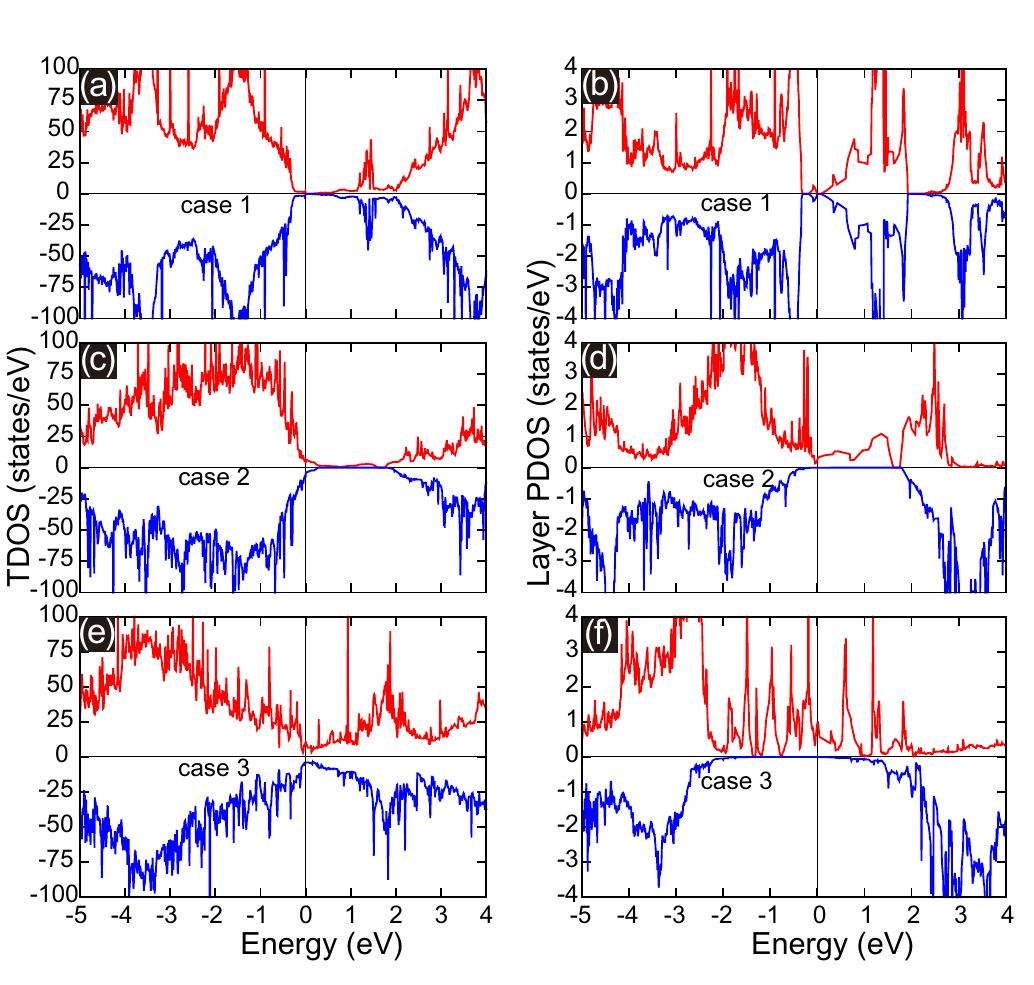}
\caption{(Color online) Total density of states (TDOS) of case-$N$ (N = 1, 2, 3) in the left panel and 
the corresponding MnO$_2$ layer projected density of states (PDOS) in the right panel. (a) TDOS of case 1.
 (b) Layer PDOS of MnO$_2$ in case 1. (c) TDOS of case 2. (d) Layer PDOS of MnO$_2$ in case 2. (e) TDOS of
  case 3. (f) Layer PDOS of MnO$_2$ in case 3. 0 eV is the reference for the Fermi level.}
\label{fig:6cap-DOS}
\end{figure}

The models of the three heterointerfaces all contain two symmetrical 6 capping layers, their relaxed structures are presented in Figure \ref{fig:models6capping}. The alignment of spin ordering of Mn cations located in a $\sqrt{2} \times \sqrt{2} \times 1$ in-plane MnO$_2$ has two possible configurations, i.e. ferromagnetic or antiferromagnetic ordering.
In order to get the magnetic ground state of monolayer MnO$_2$ in the three cases, we performed 
total energy calculations for each case. Table. \ref{tab:groundstate} shows results of respective 
ground state, the magnetic moments of Mn, and the total magnetic moment normalized to each $\sqrt{2} \times \sqrt{2} \times 1$ in-plane MnO$_2$.
 The preferred spin configuration of 
monolayer MnO$_2$ in case 1 is antiferromagnetic ordering, while Mn-Mn exchange interactions 
both energetically favor ferromagnetic coupling in the other two cases. The emergent antiferromangetic-magnetic transition in monolayer MnO$_2$ is interesting because it implies that the ground state of monolayer MnO$_2$ can be altered by its adjoining layers, which provide new freedom to control the confined properties in monolayers. The decrease of the magnetic moments of Mn, as can be seen from the third column of  Table. \ref{tab:groundstate},
 is not only the consequence of the transition of magnetic structures of monolayer MnO$_2$, 
 also indicates that the electric properties vary dramatically with the cases. Especially, 
 there are more localized moments lying around Mn sites in case 3.
\begin{table}%[!ht]
\caption{Magnetic ground state, magnetic moment of Mn ($\mu_B$$\slash$Mn) and 
total magnetic moment ($\mu_B$) in the three heterointerfaces. The total magnetic moment is normalized to a $\sqrt{2} \times \sqrt{2} \times 1$ in-plane MnO$_2$. AFM (FM) denotes antiferromagnetic (ferromagnetic) ground state.
}
\begin{center} \footnotesize
\begin{tabular}{l  c c c c c c l|c|c|c|c|} \hline \hline %
 Case name & Ground state  &  Magnetic moment$\slash$Mn ($\mu_B$) & Total Magnetic moment ($\mu_B$) \\
		\hline
 case 1 & AFM & \ \ \ \ \ \ \ \ \ \ \ \ 2.46 &  \ \ \ \ \ \ \ 0 \\
 case 2 & FM & \ \ \ \ \ \ \ \ \ \ \ \ 3.33 & \ \ \ \ \ \ \ 6.66  \\
 case 3 & FM & \ \ \ \ \ \ \ \ \ \ \ \ 4.04 &\ \ \ \ \ \ \ 8.08 \\
\hline
\hline
\end{tabular}
\end{center}
\label{tab:groundstate}
\end{table}
We now study the electronic properties of the three cases. Figure \ref{fig:6cap-DOS}(a,c,e) shows the 
total density of states (TDOS) of three heterointerfaces. It is clearly observed Fermi level is shifted
 into higher energy as the case moves from case 1 to case 2 to case 3. This shifting indicates that more 
 electronic states are transferred into monolayer MnO$_2$. Case 1 is an insulator with a narrow band gap, 
 its insulating state verifies electronic reconstruction does not occur at the interface. On the contrary, 
 bands crossing the Fermi level at the interfaces prove the existence of electronic reconstruction in case 
 2 and 3. These results are in good agreement with the results analyzed from the polar catastrophe model 
 in Figure \ref{fig:FIG1}. Apart from the system properties of the whole heterointerfaces, more inviting 
 properties are only confined in the monolayer MnO$_2$. As described by the MnO$_2$ layer projected density 
 of states (PDOS) in Figure \ref{fig:6cap-DOS}(b,d,f), the majority and minority spin electrons make equal 
 contributions to the layer PDOS with a band gap $\sim$ 0.2 eV in both spin channels, resulting from the 
 antiferromagnetic ground state in case 1. This feature of MnO$_2$ is similar to its parent phase. The 
 band gap of \SMO{} is only $\sim$ 0.15 eV if G-type antiferromagnetic structure is imposed on it based 
 on our calculations. Whereas the layer PDOSs of the remaining cases are dramatically different from that 
 in the former case. More specifically, a highly confined 100$\%$ spin-polarized 2DEG is formed in monolayer
  MnO$_2$ in both case 2 and case 3, with a minority spin band gap $\sim$ 1.8 eV for case 2 and $\sim$ 
  0.8 eV for case 3, respectively. That is to say, metal-insulator transition in monolayer MnO$_2$ is 
  realized with the accompanying of transformation of ground states.

\begin{figure}
\centering
\includegraphics[width=\linewidth]{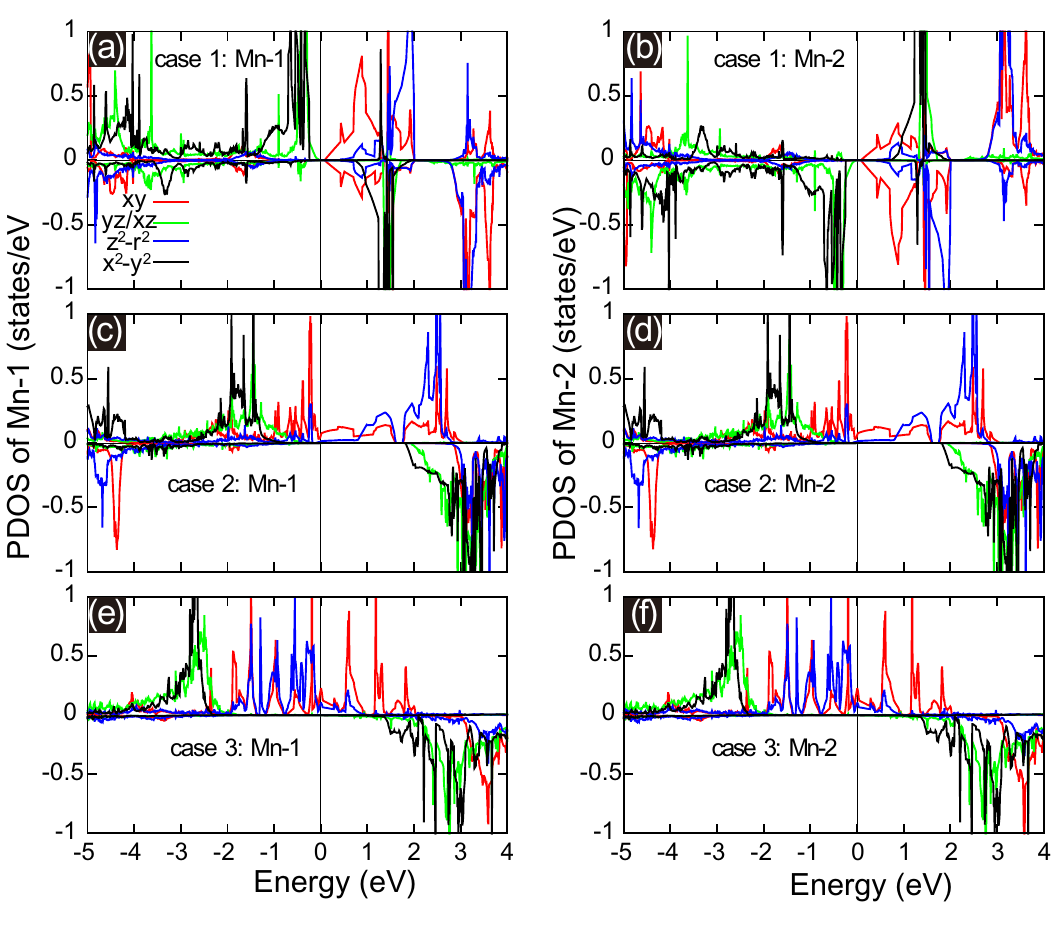}
\caption{(Color online) PDOS of Mn-1 and Mn-2 in case-$M$ (M = 1, 2, 3) with spin and orbital resolution.
 (a) PDOS of Mn-1 in case 1. (b) PDOS of Mn-2 in case 1. (c) PDOS of Mn-1 in case 2. (d) PDOS of Mn-2 in 
 case 2. (e) PDOS of Mn-1 in case  3. (f) PDOS of Mn-2 in case 3. $d_{xy}$, $d_{yz}$$\slash$$d_{xz}$, 
 $d_{z^2-r^2}$, and $d_{x^2-y^2}$ are plotted by red, yellow, blue, and black solid curves, respectively. 
 0 eV is the reference for the Fermi level.}
\label{fig:6cap-MnPDOS}
\end{figure}

A deeper understanding of the metal-insulator transition in MnO$_2$ can be gained by studying the orbital resolved PDOS. Because the models we use all contain two symmetric $\sqrt{2} \times \sqrt{2} \times 1$ 
in-plane MnO$_2$ atomic layers, each model contains four Mn atoms that can be divided into two equivalent groups
 according to the mirror symmetry of the structure. There are two atoms in each group, we will label them as 
 Mn-1 and Mn-2. Figure \ref{fig:6cap-MnPDOS} demonstrates the PDOS of Mn-1 and Mn-2 in each case. 
 Locating in the octahedral crystal field framed by the oxygen ligands, the Mn 3$d$ orbitals split into
  triplet $t_{2g}$-like ($d_{xy}$, $d_{yz}$, $d_{xz}$) and doublet $e_g$-like ($d_{x^2-y^2}$, $d_{z^2-r^2}$) states 
  to lower the energy. In case 1, Mn-1 and Mn-2 contribute equally to PDOS, but in different spin channels 
  due to the antiferromagnetic coupling between Mn-1 and Mn-2. It is the hybridizations of $d_{xy}$ with a
   gain of the strongest energy lowering and O 2$p$ states that determine the band gap ($\sim$ 0.2 eV) 
   of MnO$_2$. From -2 to 0 eV, the PDOS is mainly composed of $d_{x^2-y^2}$, $d_{yz}$ and $d_{xz}$ states. 
   More importantly, the magnetism of Mn in case 1 is mostly derived from the spin splitting of these states. 
   The majority part of $d_{xy}$ and $d_{z^2-r^2}$ states lie in much deeper energy and show relatively smaller
    spin splitting. In case 2 or 3, with parallel spin ordering, Mn-1 and Mn-2 are equivalent in symmetry, 
    hence PDOS of Mn-1 is exactly the same to that of Mn-2. The PDOS of Mn-1 in case 2 depicted in Figure
     \ref{fig:6cap-MnPDOS}c displays a metallic majority spin channel and a $\sim$ 1.8 eV band gap in 
     the minority spin channel. PDOS of Mn-1 (Figure \ref{fig:6cap-MnPDOS}e) in case 3 also shows 100$\%$ 
     spin-polarization, but the minority spin band gap is reduced to 0.8 eV. The decrease of the band gap 
     in the minority spin channel is caused by the energy shift of $d_{yz}$$\slash$$d_{xz}$ and $d_{x^2-y^2}$ 
     from case 2 to case 3. It is noticeable that $d_{xy}$ and $d_{z^2-r^2}$ states are lying in lowering 
     energy ranges in cases 2 and 3, compared to those in case 1. What is more, $d_{xy}$ and $d_{z^2-r^2}$ 
     sates are responsible for the metallicity in the majority spin channel in case 3, the respective PDOS 
     of them at the Fermi level are $\sim$ 0.25 and $\sim$ 0.125 sates$\slash$eV; while in case 2, PDOS 
     at the Fermi level is crucially dominated by $d_{xy}$ sates ($\sim$ 0.25 sates$\slash$eV at the 
     Fermi level). This discrepancy implies a much strong electronic reconstruction in case 3 because
      of the presence of sandwich-like -LaO-MnO$_2$-LaO interface as mentioned in the previous part.
Hence, we turn to study the charge state of Mn cation in each case, although qualitative results have been predicted in section~\ref{results-sec1}. A quantitative method to investigate the charge transfer is indispensable to gain a lucid understanding of the charge state of Mn cation. In general, such a quantitative prediction of charge state in large models containing about 200 atoms in first-principles calculations is usually not easy and time-consuming. 
Here, we adopt a grid-based Bader analysis algorithm\cite{Bader2006} to study the charge density of specific transition metal atoms. The calculated Bader charge of Mn cation in each case is 5.0, 5.3, and 5.46, respectively. The increase of Bader charge indicates that Mn cation in case 2 (case 3) gains 0.3 (0.46) more electrons than that in case 1. Another quantitative way is also used to confirm this result. By integrating the orbital
       resolved PDOS to the Fermi energy, we find that valence electrons of Mn cation in case 2 has 0.15
        electron less than that of case 3, and 0.21 electron more than that of case 1. One can see the 
        results derived from these two methods are consistent.

\subsection{Effects of the thickness of capping layers}
In the above parts, we have reported a systematic study of magnetic and electronic properties for the three 
heterointerfaces with six capping layers. However, there remain two questions driving this study to move 
forward. How does the thickness of capping layers affect the ground state of monolayer MnO$_2$ in case 2 
and 3? Does the spin-polarized 2DEG have a critical thickness as that in (001) and (100) \LAO/\STO{} 
interfaces\cite{Theil2006,Annadi2013}{}? To answer these two questions, we investigate the effect of 
the thickness of capping layers on the electronic and magnetic properties of monolayer MnO$_2$. Because
 spin-polarized 2DEG only exists in case 2 and case 3, we only address the issues for these two cases 
 with \LAO\ capping layers.

\begin{figure}
  \centering
  \includegraphics[width=\linewidth]{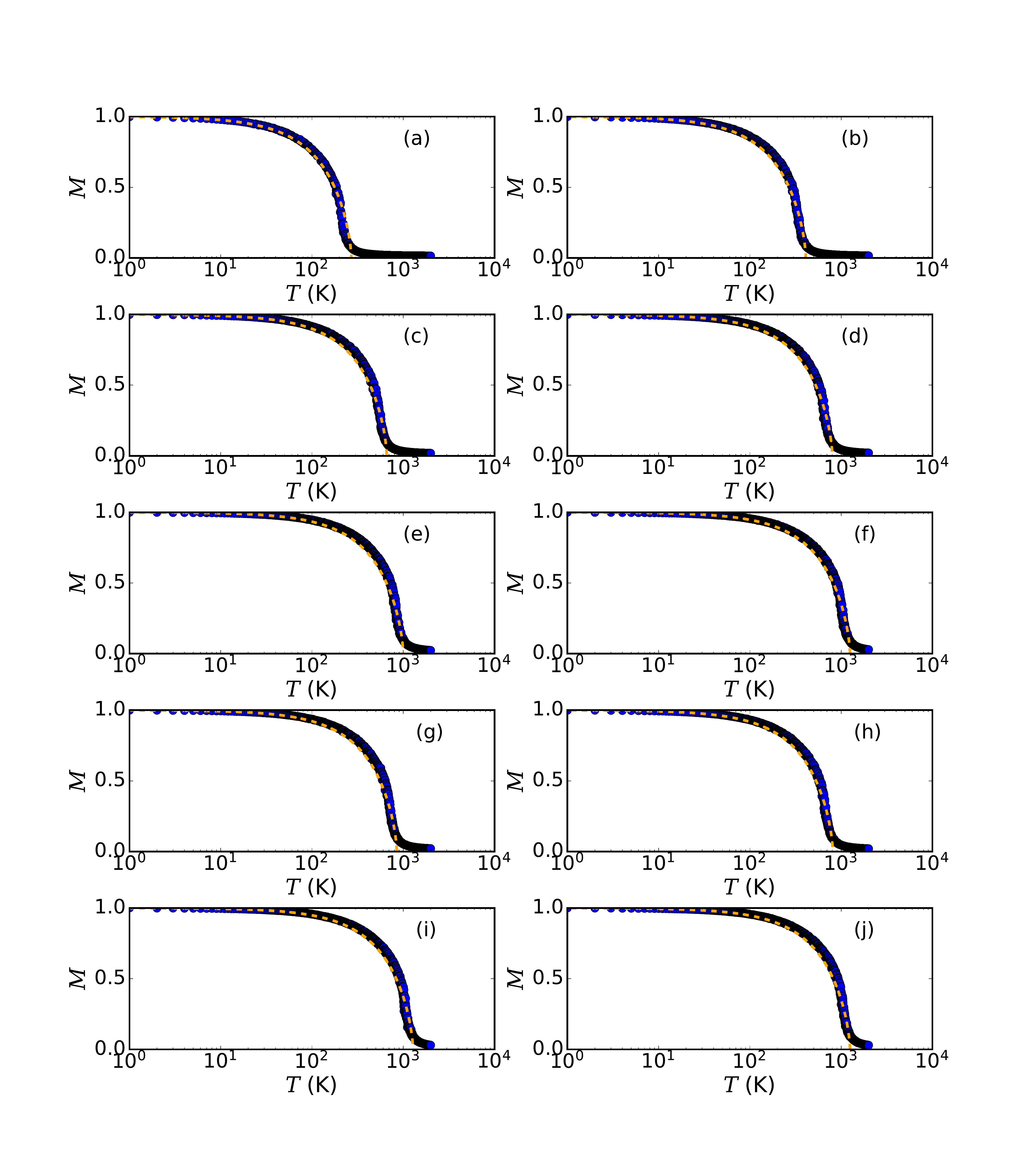}
  \caption{(Color online) 
  The normalized magnetization $M$ as a function of temperature $T$ (K). Note that the axis of temperature uses logarithmic scale. (a-e) correspond to case 2 in which capping layers $N$ = 1, 2, 3, 4, 6. (f-j) correspond to case 3 in which capping layers $N$ = 1, 2, 3, 4, 6. The circles are data from the Monte Carlo simulations, and the orange dashed lines are fitted from Curie-Bloch equation. The Monte Carlo simulations are simulated from 0 to 2000 K. 
  }
  \label{fig:monte-carlo} %
\end{figure}

% Considering the ferromagnetism in MnO$_2$ monolayer persists at room temperatures, our Monte Carlo simulations should have underestimated the TC. Such an underestimation may come from that in the actual samples, the SRO was grown at the interface in which the neighboring STO could help stabilize the ferromagnetism in SRO up to the room temperatures. 

 case 2 and case 3 with different capping layers ($N$ = 1, 2, 3, 4, 6). As can be seen, both case 2 and
  case 3 are stabilized in ferromagnetic ground states. It manifests thickness of capping layers does not 
  affect the ground states in the two cases. To estimate the Curie temperature, we further performed Monte Carlo simulations using the VAMPIRE code\cite{Evans-2014-JPCM}{} to predict the Curie temperature of the MnO$_2$ monolayer quantitatively. The Monte Carlo algorithm for classical spin models developed by Hinzke and Nowak was used\cite{Asselin-2010-PRB}. We employed a large enough thin film model with a total length of 20 nm along ${x}$ and ${y}$ directions in the Monte Carlo simulations. The simulations begin at 0 K and end at 2000 K with a temperature increment of 1 K. The equilibration time steps and loop time steps in VAMPIRE calculations are both set to 50000 to guarantee the equilibration of each step. Figure~\ref{fig:monte-carlo} shows the normalized magnetization as a function of temperature $T$ from the Monte Carlo simulations. The Curie temperature $T_C$ is estimated by fitting the data in Figure~\ref{fig:monte-carlo} with the Curie-Bloch equation in the classical limit given by ${M(T) = (1 - \frac{T}{T_C})^\beta}$~\cite{Liu2021-appliedmatertoday}.

  Our results also show, notwithstanding ferromagnetic ground states in these MnO$_2$ monolayers are energetically
     favored, magnetic moments of Mn cation show thickness-dependent behaviors. Specifically, the magnetic 
     moment of Mn cation in case 2 decreases when \LAO\ layers become thinner; the situation in case 3 is a
      bit different, as \LAO\ thickness drops from $N$ = 6 to $N$ = 2, the magnetic moment of Mn cation is
      decreased from 4.04 $\mu_B$ for $N$ = 6  to 3.78 $\mu_B$ for $N$ = 2, yet there is a modest increase
       if one more capping layer is removed (i.e. $N$ = 1), the magnetic moment becomes 3.86 $\mu_B$.

\begin{table*}[!ht]
 \caption{
 The magnetic ground state, magnetic moment of Mn ($\mu_B$), and calculated Curie temperature from Monte Carlo simulations for case 2 
 and case 3 with varying thicknesses of capping layers. FM denotes the ferromagnetic ground state.}
\begin{center} \footnotesize
\begin{tabular}{l  c c c c c c l|c|c|c|c|} \hline \hline %
Case name & Capping layers & Ground state  & Magnetic moment$\slash$Mn ($\mu_B$)  & $T_c$ (K) \\
		\hline
case 2 & 6 & \ \ \ \ \ FM & \ \ \ \ \ \ \ 3.33 & 1000 \\
case 2 & 4 & \ \ \ \ \ FM & \ \ \ \ \ \ \ 3.32 & 810 \\
case 2 & 3 & \ \ \ \ \ FM & \ \ \ \ \ \ \ 3.20 & 660 \\
case 2 & 2 & \ \ \ \ \ FM & \ \ \ \ \ \ \ 3.18 & 410 \\
case 2 & 1 & \ \ \ \ \ FM & \ \ \ \ \ \ \ 3.11 & 270 \\
case 3 & 6 & \ \ \ \ \ FM & \ \ \ \ \ \ \ 4.04 & 1250 \\
case 3 & 4 & \ \ \ \ \ FM & \ \ \ \ \ \ \ 4.01 & 1280 \\
case 3 & 3 & \ \ \ \ \ FM & \ \ \ \ \ \ \ 3.83 & 810 \\
case 3 & 2 & \ \ \ \ \ FM & \ \ \ \ \ \ \ 3.78 & 850 \\
case 3 & 1 & \ \ \ \ \ FM & \ \ \ \ \ \ \ 3.86 & 1250 \\
\hline
\hline
\end{tabular}
\end{center}
\label{tab:ground-thick}
\end{table*}

\begin{figure}
\centering
\includegraphics[width=.98\linewidth]{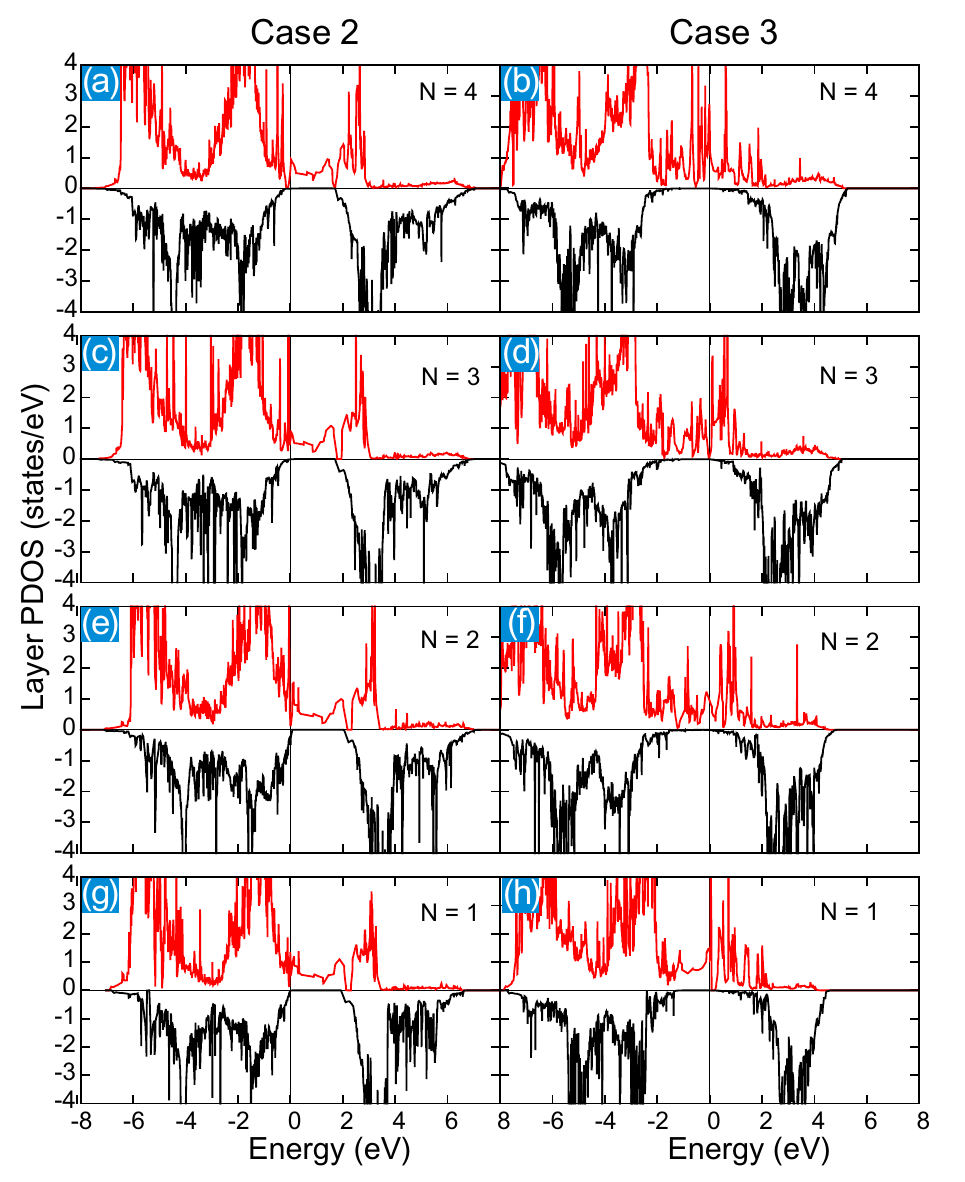}
\caption{(Color online) MnO$_2$ layer projected PDOS. Left panel: PDOS of case 2 with $N$ capping layers 
[$N$ = 4 (a), 3 (c), 2(e), and 1(g)]. Right panel: PDOS of case 3 with $N$ capping layers [$N$ = 4 (b), 3
 (d), 2(f), and 1(h)]. 0 eV is the reference for the Fermi level. Red and yellow solid curves denote 
 majority and minority spin channels, respectively}
\label{fig:thicknessDOS}
\end{figure}

\emph{Electronic properties.} Figure \ref{fig:thicknessDOS} displays the MnO$_2$ layer PDOS in case 2 and case 
3 with thickness dependence. Except for the abrupt decline in spin polarization ($\sim $23.0$\%$) of MnO$_2$ 
in case 2 with $N$ = 2, the remaining MnO$_2$ layers all show well-defined half-metallicity. For special 
case 2 with two capping layers, we find the orbital occupation of O $p_z$ in the minority spin channel is responsible 
for the remarkable reduction of spin polarization of monolayer MnO$_2$. Furthermore, the exchange splitting of O 
$p_z$ introduces $\sim$ 0.05 $\mu_B$ local moments at the sites of oxygen.

\begin{figure}
\centering
\includegraphics[width=.95\linewidth]{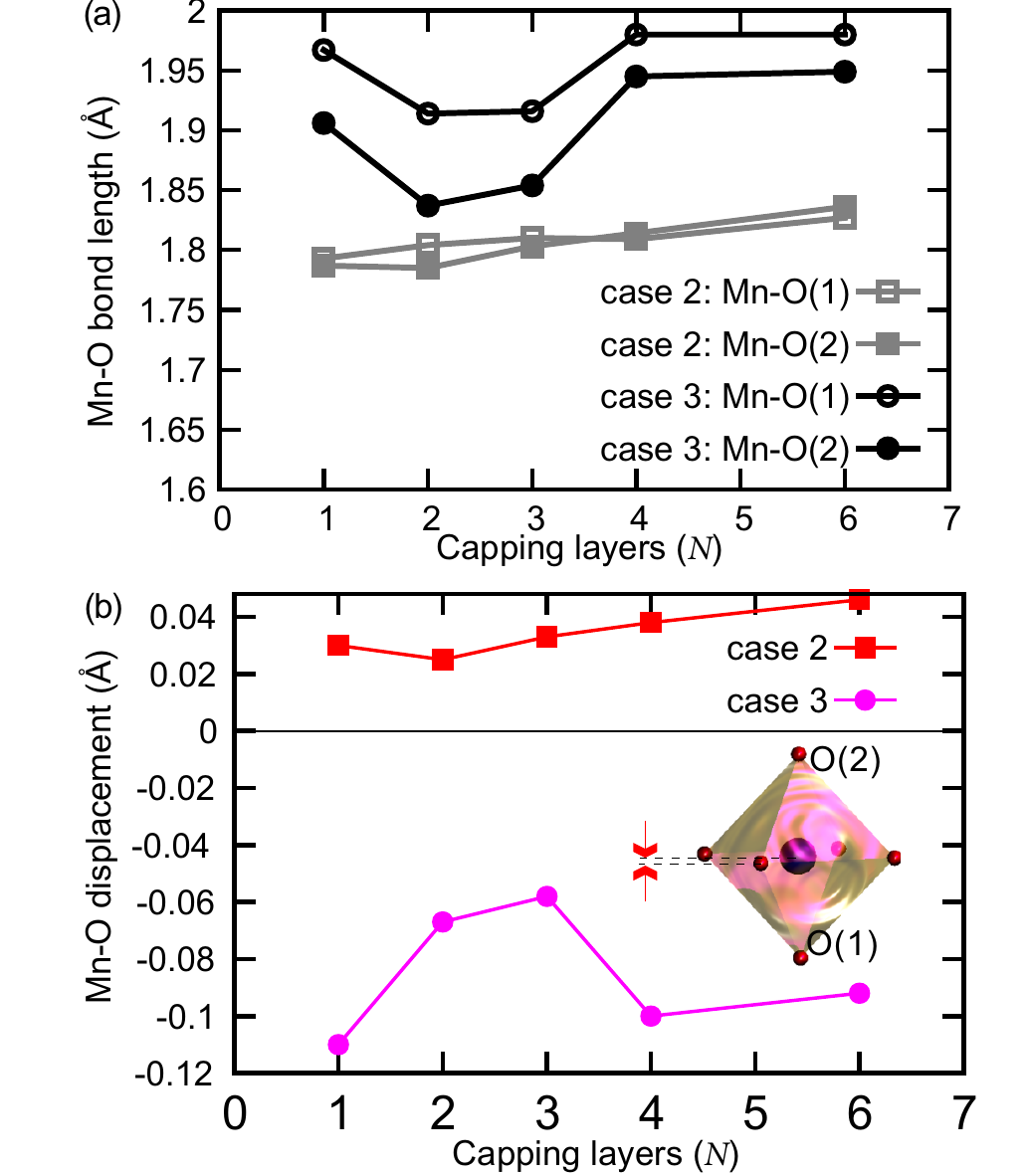}
\caption{(Color online) Mn-O displacement and Mn-O bond length against the thickness of capping layers. (a) 
The bond lengths of Mn-O(1) and Mn-O(2) as the function of the thickness of capping layers. Here, we assign 
O(1) in the substrate and O(2) in the capping layer. (b) Mn-O displacements (\AA) along $z$-direction as
 the function of the thickness of capping layers. The spacing indicated by the two red arrows illustrates
  the Mn-O displacement in the octahedral MnO6 cage. Positive (negative) sign in the displacement axis
   means $z_{\rm Mn}$ is bigger (smaller) than $z_{\rm O}$}
\label{fig:relatingstruct}
\end{figure}

\emph{Structure interpretation.} In order to build a relationship between the structure and the above 
observations qualitatively. We examined the Mn-O(1)/Mn-O(2) bond length and Mn-O displacements 
perpendicular to the plane of the MnO$_2$ monolayer. The left panel of Figure \ref{fig:relatingstruct} 
presents Mn-O(1)/Mn-O(2) bond length as a function of the thickness of capping layers. We note that 
the bond length of Mn-O(1)/Mn-O(2) in case 2 is decreased with the decreasing thickness, whose trend 
is the same as that of the magnetic moment of Mn ions in this case. In case 3, the bond length of Mn-O is 
decreased with the reducing capping layers when $N$ $\ge$ 2, but it is increased when $N$ = 1, this 
result coincides to the trend of magnetic moment in case 3. Therefore, the magnetic moments of Mn cations 
are fundamentally associated with the bond length of Mn-O. In the right panel of Figure \ref{fig:relatingstruct}, 
we also analyze the Mn-O displacements so as to interpret the introduced 0.05 $\mu_B$ magnetic moment at the oxygen 
site. By surveying all the data in the right panel, it is found the Mn-O displacement in case 2 with $N$ = 2 is
 the smallest ($\sim$ 0.025 \AA) among all the calculated structures, in this situation, Mn-O covalency bonding is
  stronger than other structures by hybridizing Mn-3$d$ with O-2$p$ orbitals.

\section{Discussion}
By designing three heterointerfaces with varying thicknesses of capping layers, metal-insulator transition, 
as well as the magnetic transition was revealed in monolayer MnO$_2$ by analyzing the electronic and magnetic 
properties. The metal-insulator transition and ferromagnetic-antiferromagnetic transition are independent on
 the thickness of the capping layers but are linked to the adjoining layers that determine the occurrence
  of electronic reconstruction. The vanishing critical thickness of the 100$\%$ spin-polarized 2DEG is very
   different from that in \LAO$\slash$\STO\ with the critical thickness of three layers\cite{Hwang.NAT.2004}{}. 
   We think this discrepancy is chiefly ascribed to the band gap of monolayer MnO$_2$ and in-built electric field in the capping layers. The band gap of monolayer MnO$_2$ is only about 0.2 eV in calculations, which can be easily destroyed by the 
   sizable internal electric field $\sim$0.24 V$\slash$\AA{} in a unit-cell \LAO~\cite{LeeJaekwang2008PRB}. As a result, critical thickness vanishes.

Our current findings show the magnetic transition is closely correlated to the metal-insulator transition.
 In case 1, we find the 180$^\circ$ Mn-O-Mn bond is well preserved. Being an insulator containing $t_{2g}^3$$e_{g}^0$ magnetic ions, its spin ordering can be understood by the Goodenough-Kanamori-Anderson (GKA) rules\cite{Kanamori.1959}. A 180$^\circ$ Mn-O-Mn bond can give rise to antiferromagnetic superexchange by hopping between two half-filled active orbitals and the same ligand $p$ orbital (i.e. $pd\pi$ hopping in this case)~\cite{Sun2017-BFMO}. Differing from case 1, case 2 and case 3 display spin-polarized conductivity. In these two systems, the magnetic interactions are dominated by the coupling between the localized and itinerant electrons. This exchange, also known as double exchange or Zener exchange\cite{Zener}, leads to the ferromagnetic coupling, and has also been reported to stabilize the ferromagnetic order in the (\LMO)$_n$$\slash$(\STO)$_m$ superlattices\cite{Jilili-Udo2015} and \LMO$\slash$\STO{} heterostructures with hole or electron doping\cite{Tsymbal2015}{} recently.

\section{Conclusion}
In summary, monolayer MnO$_2$ is designed to be sandwiched at complex oxide heterointerfaces, 
their electronic and magnetic structures with varying thicknesses of capping layers are systematically 
investigated by the first-principles calculations. Due to the presence of charge reconstruction at the 
interface, metal-insulator transition and magnetic transition are both observed in the one-atom-thick
 MnO$_2$. Our results show these transitions are independent of the thickness, which makes it stand out
  from (001) and (110) \LAOSTO{} with critical thickness. In addition, 100$\%$ spin-polarized 2DEG with
   robust room-temperature ferromagnetism are reported in the buried monolayer MnO$_2$, we predict it can 
   potentially be applied to the next-generation nanoscale spintronic devices. This work presents fascinating
    phenomena in monolayer transition metal with first-principles calculations and paves an innovative way to
     design novel two-dimensional materials. We expect the present study will motivate more efforts to study
      the two-dimensional transition metal oxides and their future applications in various electronic$\slash$spintronic devices.

% \section{Authorship contribution statement}
% Rui-Qi Wang: Investigation,  Analyzing, Writing $\&$ Review.
% Tian-Min Lei: Funding acquisition $\&$ Review.
% Yue-Wen Fang: Supervision, Investigation, Interpretation, Writing $\&$ Review.

% \section*{Acknowledgements}
% The authors acknowledge the computational resources provided by the ECNU computing center during their stay at ECNU.

\section*{Acknowledgements}
The authors R.-Q.W. and Y.-W.F. acknowledge the discussions with Chun-Gang Duan during their stay at East China Normal University (ECNU). Y.-W.F. thanks the support from his current laboratory led by Ion Errea for the great support on the research activities. The computations were primarily performed by Y.-W.F. at the High-Performance Computing Center of ECNU, with minor contributions from R.-Q.W. who is supported by the Scientific Research Program Funded by Shaanxi Provincial Education  Department (Program No.21JK0699) and School level fund of Xi'an Aeronautical Institute (Program No.2020KY1224). Y.-W.F. supervised and investigated this project, and wrote the manuscript during his stay at ECNU. R.-Q.W. improved the manuscript by rewriting the introduction, interpreting and appending computational results on interface stability with T.-M.L.'s assistance. All authors have contributed to reviewing the manuscript.

\section*{Declaration of competing interest}
The authors declare no competing interests.

\section*{Data availability}
All data that support the findings of this study are included in the article, and the freely open database Zenodo~\cite{fang_2023_zenodo-MnO2}.

% \bibliographystyle{iopart-num-ywfang}
% \bibliography{reference-MnO2}

 \end{document}